\documentclass[prb,showpacs,twocolumn]{revtex4}
\usepackage{graphicx}
\usepackage{dcolumn}
\usepackage{bm}

\begin{document}
\title{Solvation of positive ions in water: The dominant role of water-water interaction}
\author{Christian Krekeler}
\author{Luigi Delle Site}
\affiliation{%
Max-Planck-Institut f\"ur Polymerforschung, Ackermannweg 10, D-55128 Mainz, Germany
}%

\begin{abstract}
Local polarization effects, induced by mono and divalent positive ions in water, influence (and in turn are influenced by) the large scale structural properties of the solvent. Experiments can only distinguish this process of interplay in a generic qualitative way. Instead, first principles quantum calculations can address the question at both electronic and atomistic scale, accounting for electronic polarization as well as geometrical conformations. For this reason we study the extension of the scales' interconnection by means of first principle Car-Parrinello molecular dynamics applied to systems of different size. In this way we identify the general aspects dominating the physics of the first solvation shell and their connection to the effects related to the formation of the outer shells and eventually the bulk. We show that while the influence of the ions is extended to the first shell only, the water-water interaction is instead playing a dominant role even within the first  shell independently from the size or the charge of the ion.
\end{abstract}
\maketitle
\section{introduction}
Water is the natural solvent in numerous processes occurring in day life phenomena; from biophysics to industrial applications the role of water as a solvent is crucial to understand the physics and chemistry behind a process.\cite{applic} For this reason, such systems are among those most studied both experimentally\cite{jalili2001} and theoretically.\cite{feller1993,adrian2005,dudev1999,alcami1999,pavlov1998,katz1996,bock1993,mueller2004,naor2003,bako2002,lightstone2001,lightstone2005,whitegalli2000,lyub2001,hribar2002,ramaniah1999,ikeda2007,carrillo2003,blake1995} Nevertheless many questions remain still open and require novel approaches that are able to reveal those aspects which are neither captured by the experiments, because of the technical limitation of the accuracy, nor by theoretical models, because of the inherent scale limitations of the specific theoretical approach. 
The purpose of this work is to conduct a systematic study of the ion solvation comparing positive ions of different size and charge. In particular the analysis is focused on the characterization of the electronic polarization and structural reorganization of the solvent due to the presence of the ion.  
For this purpose we use first principle quantum calculations, performed via the Car-Parrinello Molecular Dynamics code (CPMD) and study the solvation of different positive ions, namely the monovalent $Li^{+}$, $K^{+}$, $Na^{+}$, and the divalent $Ca^{2+}$ and $Mg^{2+}$.
The strategy employed consists of considering for each ion, systems of different sizes, that is from one water molecule to the cluster which corresponds to the coordination number of the ion and finally the study is extended to large systems, prototypes of a realistic solvation. This stepwise analysis allows for an accurate detection of the changes in the electronic polarization produced by each additional water molecule in the cluster; furthermore, it allows to clarify, by comparison with the large systems, the role played by the bulk into the solvation process; this can be interpreted as the interplay between the local scale of ion-induced electronic polarization and the global scale of structural reorganization. In this sense we introduce, w.r.t. previous work a different point of view. In fact previous studies, either theoretical \cite{naor2003, bako2002, lightstone2001, lightstone2005, whitegalli2000, lyub2001, hribar2002, ramaniah1999, ikeda2007,carrillo2003,blake1995} or experimental \cite{jalili2001}, focus mainly on the determination of the coordination number of ions and on their dynamic evolution. The molecular polarization, not detectable by experiments, is also quantified by first principles quantum calculation for the first solvation shell and bulk \cite{naor2003,bako2002,lightstone2001,lightstone2005,whitegalli2000,lyub2001,ramaniah1999,ikeda2007}. However such a work does not focus on the understanding of the scales' interplay by  comparison between ions of different size and charge. For this reason this study represents a complementary view which, together with the past work, will allow for a better understanding of water solvation. 
The starting point of our analysis consists in drawing a generic picture of the solvation as follows: the positive charge of the ions induces an electronic polarization of the closest water molecules and their dipole tends to align along the oxygen-ion direction. At the same time water molecules experience their mutual presence which leads to the competing effect of building the hydrogen bond network with the neighboring molecules; this happens around the ion as well as far from it. Having outlined the main ingredients of the process, the crucial question is how they are linked to each other. The answer we found in terms of electronic polarization is interesting; we identify in the water-water interaction a dominating effect which  drives most of the electronic polarization, reducing the ion influence to merely geometrical properties, localized in the first shell, due to the structural rearrangement of the hydrogen bond network. This is caused by the fact that the molecules are hindered to occupy the excluded volume of the ion. This scenario emerges by analyzing the molecular dipole moment and its orientation using the Wannier decomposition.\cite{marzri1997,souza2001,resta1998} In this sense, only static properties are considered, although for the large systems (i.e. 32,64 and 128) equilibration is achieved before data are collected for analysis. 
\section{Technical Details}      
We employ the Car-Parrinello, plane-waves Density Functional based code CPMD \cite{cpmd392} to study the hydration of three monovalent ions of different size, respectively, $Li^{+}$, $Na^{+}$, $K^{+}$ and divalent ions, $Mg^{2+}$, and $Ca^{2+}$. The systems considered are small clusters, from one water molecule to the cluster containing a number of molecules equal to the coordination number; for each cluster several geometrical arrangements were considered and geometry optimization was carried until the maximum component of the force was below the threshold of $10^{-3}$a.u.. For the monovalent ions, these were taken from our previous work \cite{krek2006}. Next, large systems were considered 
which contain respectively 32, 64 and 128 molecules. 
For each of these systems a simulation was carried out. For 32 molecule the lateral size of the simulation box is $9.5 \AA$ the electron mass was chosen 400 a.u. and the time step 3 a.u.. 
The systems were equilibrated for 5 ps and the data were collected during the next 5 ps of simulation. The same was done for the systems of 64 molecules and in this case the lateral cell dimension was $12\AA$ and the same parameters for the electron mass and the time step were used. For the 128 molecules systems, being used only as a further check, due to the highly demanding computational cost, the equilibration was done for 3 ps and data collected for the next 2 ps; in this case the lateral size of the box was $15\AA$ the electron mass was chosen to be 600 a.u. and the time step 4 a.u..
In any case we checked that the equilibration was sufficient by comparing the various radial distribution functions with those obtained for the 64 system and those available in literature. For all calculations the BLYP gradient correction was used \cite{perdew1986,becke1986,lyp1987}. For all ions but $Li^{+}$ we employed Troullier-Martin pseudopotentials which include the semicore electrons, while for $Li^{+}$ a non linear corrected pseudopotential was used. All pseudopotentials  were accurately tested to reproduce reference data \cite{krek2006,Herzberg1966,herzberg2005}. The Maximally localized Wannier scheme \cite{marzri1997,souza2001,resta1998}, implemented in CPMD was used to determine the molecular dipole of water. 
\section{Results and Discussion}
\begin{figure}[!ht]
\includegraphics[width=7.0cm]{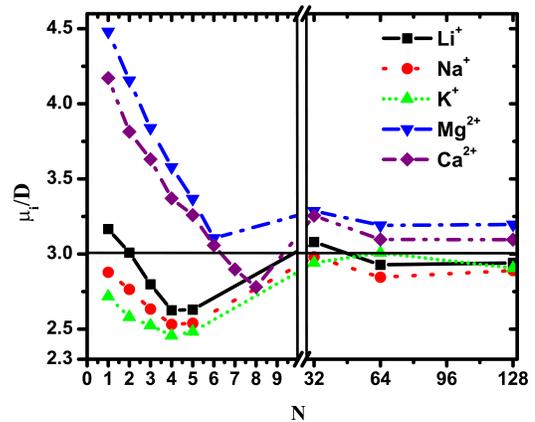}
\caption{\label{fig1} (color online) Average molecular dipole for different systems;
shown are clusters of increasing size, from a single molecule up to the characteristic coordination number of the ion considered (left), and to the  first solvation shell of extended systems of different size, 32, 64 and 128 molecules (right). To be noticed that for the case of full polarization, i.e. one water system, a strong ion dependency of the molecular dipole is found while for the case of extended systems the value is close to that of pure water with a small dependency on the specific ion.}
\end{figure}

\begin{figure}[!ht]
\includegraphics[width=7.0cm]{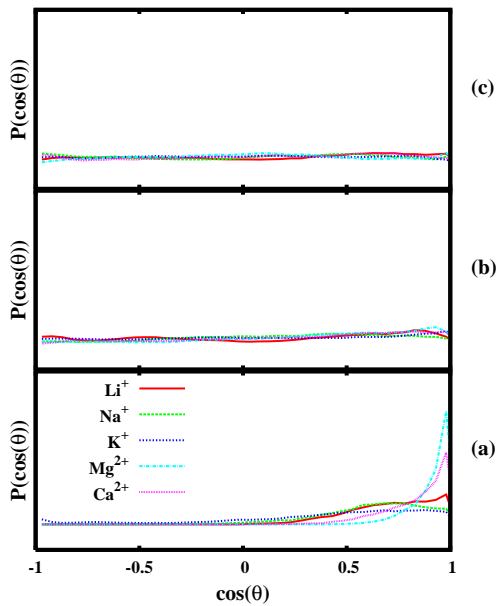}
\caption{\label{fig2} (color online) Dipole orientation with respect to the oxygen-ion direction for the case of 64 water molecules: (a) only molecules in the first solvation shell are counted; (b) only molecules in the second solvation are counted; (c) only molecules in the third solvation are counted.   
To be noticed that while in the first shell, local structures are ion dependent, beyond this shell this dependency vanishes in all cases.}
\end{figure}

\begin{figure}[!ht]
\includegraphics[width=7.0cm]{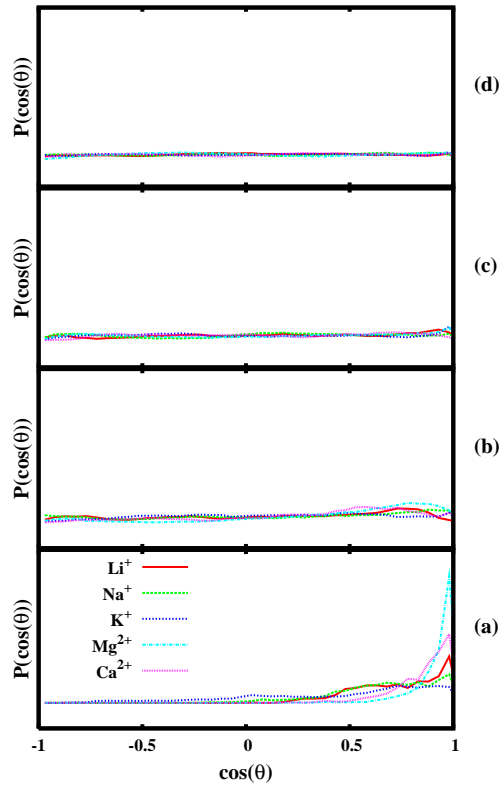}
\caption{\label{fig3} (color online) As the previous figure for 128 molecules. In this case in (d) the distribution reported is that considering the molecules of the bulk, i.e. beyond the third solvation shell.}
\end{figure}

The molecular dipole is a quantity which can directly expresses the electronic polarization due to the interaction of the molecule with the surrounding environment. For this reason we focus on the determination of such a quantity for the various systems we study. As underlined before, the advantage of building stepwise the hydration shell is that one can detect the effects on the polarization of a molecule due to the addition of another molecule into the cluster. In Fig.\ref{fig1} is reported the average water molecular dipole as a function of the number of molecules surrounding the various ions and the same quantity for the first solvation shell for the larger systems of 32, 64 and 128 water molecules.
 A general trend emerges, which was already noticed in our previous work \cite{krek2006}: by adding molecules  up to the number corresponding to the coordination of the specific ion, the value of average dipole decreases. One could think of it as the induced dipole being diminished once the field of the second water plus the field of the ion is reduced. So far we have interpreted the results regarding the cluster calculations where the role of water-water interaction is identified as an important one but still competing with the direct polarization due to the ion. To understand the real importance of water-water interaction for the solvation process we have to analyze the second part of Fig.\ref{fig1} where the effect of further solvation shells and bulk are present. In fact despite the size and the charge of the ion, the dipole in the first solvation shell is comprised between 2.9 and 3.2 Debye, which means very close to the value for pure water of about 3.0 Debye \cite{silvestrelli1999,silvestrelli1999a}.  Remarkably, if we take the two extreme curves, that is $Mg^{2+}$ and $K^{+}$, we can clearly see that the full ionic polarization in case of one molecule differs by about 2.0 Debye, however upon full hydration the difference is about 0.3 Debye. These results strongly support the picture that water-water interaction dominates the polarization process while the different ability of specific ions to polarize the molecules plays a less relevant role. As one can see from Fig.\ref{fig1} for large systems, in the case of divalent ions, the molecules in the first solvation shell have a dipole slightly larger than the one found in bulk water, while in case of monovalent ions the value is slightly smaller. This means that there is some ion-specific effect contributing to the water molecular dipole, however this is indeed small compared to the water-water polarization contribution. 
Of course the analysis above covers the quantification of molecular charge deformation, but there is another aspect that we must consider, that is the structural arrangement due to the presence of the ion. In order to solvate the ion, water must create a cavity and adapt the surrounding structure accordingly; also, due to the presence of a positive charge, the molecular dipole tends to align along the oxygen-ion line. This happens for all ions in the first solvation shell, as shown in Figs.\ref{fig2}, \ref{fig3}, but the second solvation shell shows evident less order and is the same as that of the further shells. Remarkably, while the amount of order of the first shell depends on the specific nature of the ion (charge and size) for further shells the situation is the same for all ions.
This means that the role of the ion regarding the structural arrangement of water is more important than in the case of charge polarization, as we see some preferential direction of orientation in the first shell, however its effect is very local. The overall picture emerging from this study is that the molecular polarization during solvation is determined by the water-water interaction, this latter happening on large scale via the formation of the hydrogen bond network. The role of the ion as a source of molecular polarization, expressed in term of induced deformation of the molecular charge, is secondary. On the other hand, the structural properties of water induced by the excluded volume of the ions and by its charge as a promoter of a orientational preference leads to a well determined local solvation structure which rapidly decays after the second solvation shell.
\section{Conclusions}
The results reported in this work are important for several reasons. The understanding of the dominant effects regarding the molecular polarization in solvation and its general character independent from the specific ion cannot be detected by experiments and by theoretical studies of specific systems. In this sense we contribute to a deeper understanding of what really matters in solvation of positive ions which is crucial to understand more complex systems of biological or industrial interest and to characterize once more the peculiarity of water and its hydrogen bond network. In particular, for molecular modeling, the message we give is of great utility and supports the use of standard classical models of water without additional polarization effects when in contact with an ion. Implicitly these results give indications of why classical models do in general rather well. This overall picture was already conjectured in previous work \cite{krek2006,molphys,molsim1,molsim2} and finds now a solid basis.\\
\\
{\bf Acknowledgments}
We thank B.Hess for for preparing the
classical configurations used as input in CPMD, M.Praprotnik, L.M. Ghiringhelli, N.van der Vegt and R.M.Lynden-Bell for a critical reading of the manuscript.

\end{document}